# The Blow-off Impulse Equivalence between Multi-Energy Composite Spectrum Electron Beam and Powerful Pulsed X-ray

D. W. Wang,[a,b] S. H. Yang,[a] S. Wang,[a,1] J. Wang[a] and H.P. Li[a]

[a] *Xi'an Jiaotong University,*
  *No.28, Xianning West Road, Xi'an, Shaanxi, 710049, P.R. China*

[b] *Lab of Intense Dynamic Loading and Effect,*
  *No.28, Pingyu Road, Xi'an, Shaanxi, 710024, P.R. China*
  *E-mail*: shengwang@mail.xjtu.edu.cn

ABSTRACT: The electron beam, one of the most effective approaches to simulate the irradiation effects of powerful pulsed X-ray in the laboratory, plays an important role in the experiment of simulating thermodynamic effects of powerful pulsed X-ray. This paper studies the thermodynamics equivalence between multi-energy composite spectrum electron beam and blackbody spectrum X-ray, which is helpful to quickly determine the experimental parameters in the simulation experiment. The experimental data of electron beam is extrapolated by the numerical calculation, to increase the range of energy flux. Through calculating the blow-off impulse of blackbody spectrum X-ray irradiation, we obtained the curve of X-ray blow-off impulse varying with energy flux, and then found two categories of equivalent relations - equal-energy flux and equal-impulse - by analysing the calculation results of electron beam and X-ray blow-off impulse. Based on such relations, we could directly or indirectly obtain the results of blackbody spectrum X-ray irradiation blow-off impulse via electron beam experiment.

**Contents**

## 1. Introduction

Powerful pulsed X-ray is mainly soft X-ray, featuring high energy flux and short duration (about 100ns) [1,2]. Under the irradiation of powerful pulsed X-ray, the materials within the optical thickness on the side under radiation will melt rapidly, vaporize or even partially dissociate into plasma, and blow off against the light at a high speed. The generated blow-off impulse can cause the buckling deformation and vibration of the structure, leading to its instability [3]. At the same time, due to the inhomogeneous X-ray energy deposition, inside the material will emerge thermal shock waves whose propagation and reflection can cause cylindrical shell damage and spallation damage [4]. These problems caused by powerful pulsed X-ray irradiation are collectively referred to as thermodynamic effects. Therefore, studying the thermodynamic effects of powerful pulsed X-ray is of great significance to assess the survivability of spacecraft and test the effectiveness of anti-nuclear reinforcement measures [5].

The thermodynamic effects of material and structure can be induced by powerful pulsed X-ray [6], electron beams [7], laser beams [8] and chemical explosion [9,10]. However, although the wavelength of the high-energy laser is usually in the infrared band and much longer than the X-ray, its optical thickness is usually close to zero for opaque medium. Therefore, laser is usually treated as a heat flow boundary condition in the study of laser-induced hard damage. The thermodynamic effects of powerful pulsed X-rays are studied indirectly by means of a low-energy intense-current pulsed electron accelerator [11,12]. Compared with powerful pulsed X-ray, the optical thickness of electron beam is relatively large, and the peak and gradient of energy deposition profile are relatively small. Therefore, it is necessary to further study the equivalence of electron beams and powerful pulsed X-ray from the aspects of the energy spectrum and irradiation mode. For mono-energetic electron beams, Yang [13] studied the energy deposition profile of three mono-energetic electron beams in aluminium alloy materials at different angles of incidence, and found that the irradiation effects of powerful pulsed X-rays could be well simulated when the low-energy electron beams had a larger angle of incidence on the target material.

To sum up, electron beam is an effective method to simulate powerful pulsed X-ray. However, it is necessary to study the equivalence of the above two for the thermodynamic effects caused by the radiation mechanism difference between them. When the energy flux of X-ray radiation is high enough, the vaporization and melting of the material on the light front and the outward ejection of high-temperature and high-pressure material would exert a recoil impulse (blow-off impulse) on the structure. On the one hand, the blow-off impulse exerts a compression wave (which belongs to the material response) to the material (in this case, the exposed surface will not produce sparse tensile wave due to the suppression of the light facing surface). On the other hand, the blow-off impulse may vibrate spatial structures, thus leading to permanent deformation and even buckling. Therefore, the blow-off impulse is one of the key parameters of X-ray radiation thermodynamics. In this paper, the study on the equivalence of electron beam and X-ray, or the study on the equivalent relationship between multi-energy composite spectrum electron beam and blackbody spectrum X-ray in terms of thermodynamic effects, could provide a reference for rapid determination of experimental parameters for simulation experiment.

## 2. Experimental System and Numerical Method

### 2.1 Platform for Electron Beam Experiment

The pulsed electron accelerator is used to simulate X-ray's thermodynamic effects. The analysis of the radiation loading characteristics of the Relativistic Electron Beam (REB) is shown as below. The energy ($E_0$) and the average energy ($E_{av}$) are 0.1MeV~1.2MeV and 0.5MeV~0.6MeV respectively, with the Full Width at Half Maximum (FWHM) of 50ns~80ns. When X-ray energy flux is less than 400J/cm$^2$, the beam spot diameter is 90mm~170mm on the target. By adjusting the distance between the target and the electron beam cathode, the energy flux can reach 400J/cm$^2$ ~1,000J/cm$^2$, and the beam spot diameter is 50mm~60mm.

### 2.2 Blow-off Impulse Measurement Device

Fig. 1 shows the photo and schematic structure of the blow-off impulse probe. The parallel target is attached to one end of the signalling rod whose other end is shaped into several evenly spaced rings. Moreover, infrared luminous diodes and photoelectric triodes are installed on both sides of the rod. With the rod moving forward, the rings block off the light and cut off the signal, after that, the signal is restored. The digital oscilloscope records the time interval "$\Delta t$" while the rod moves through the spacing $L$ between two rings (L=5mm). Thus, the average velocity of the rod is obtained:

$$v = L/[\Delta t(1-\mu)] \qquad (1)$$

where $\mu$ is the friction adjustment factor (4.5%), which is calibrated by a gas gun. According to the law of momentum conservation, the blow-off impulse I can be expressed by:

$$I = (m_0 - \Delta m)v/A \qquad (2)$$

where $\Delta m$ is the mass loss of the target after radiation, $m_0$ is the initial total mass of the target and the rod, and $A$ is the irradiated area ($A \approx 2.01 \text{cm}^2$). The uncertain impulse is less than 10%. If the energy flux $E_\Phi$ of soft X-ray radiated on the target is given, the coupling coefficient of blow-off impulse will be determined:

$$\beta = I / E_\phi \quad (3)$$

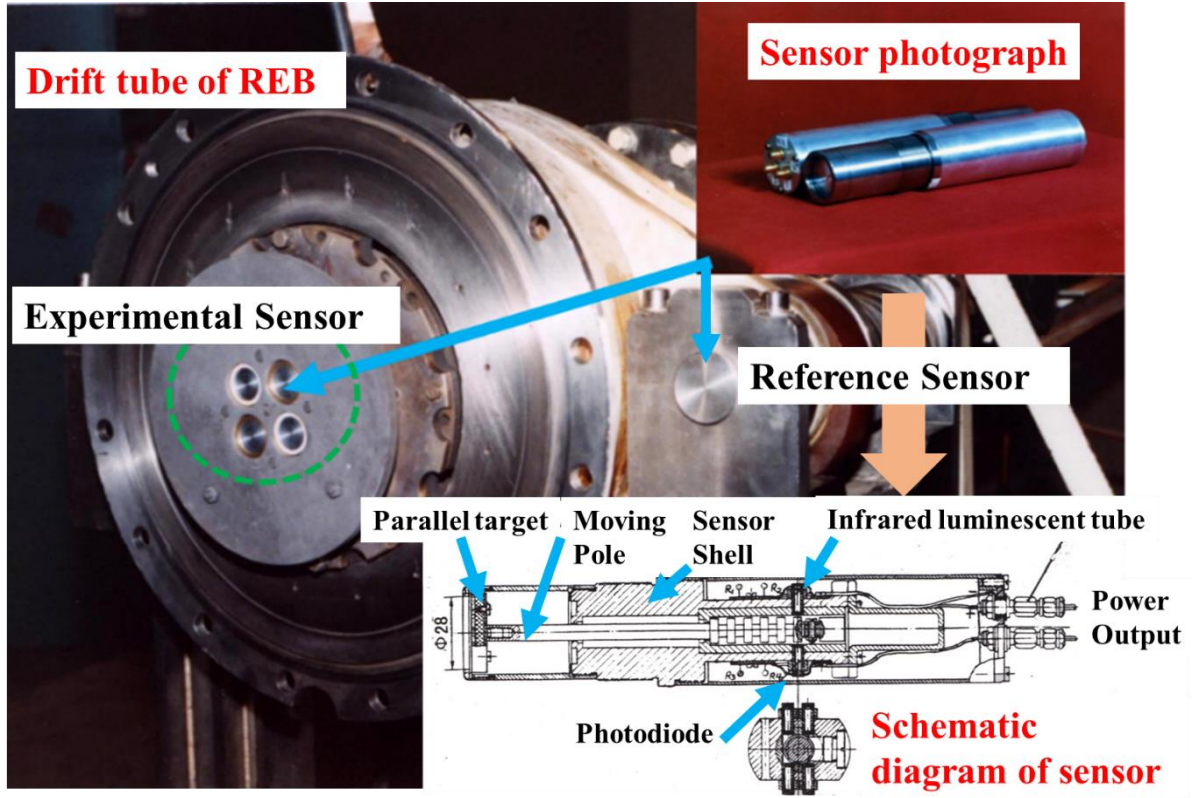

Fig. 1 Impulse sensors installed in the drift tube of REB

### 2.3 Finite Element Calculation Software

Due to the failure of the existing finite element analysis software to deal with the thermodynamic problems caused by electron beam and pulsed X-ray irradiation, we used the finite element method and the FORTRAN language to compile the program RAMA [8] for the thermodynamic effects of pulsed beam irradiation to carry out relevant numerical simulation work. Embedding a variety of constitutive models and equations of state including orthotropic dynamic elastoplastic constitutive model, the RAMA program can not only deal with the fracture and vaporization of materials, but simulate the stress wave propagation in the collision of two-dimensional flat plates by anisotropic and isotropic materials, and the thermodynamic effects of two-dimensional X-ray in various shapes. It thus has a certain practicality in engineering applications.

The calculation process of electron beam energy deposition can be summarized as follows: In a unit mass thickness material $S(E)$, the electron energy loss, mainly caused by electron collision ionization for lower-energy electrons, is as follows [16].

$$S(\text{E}) = \frac{2\pi e^2 N_0 Z}{m_0 c^2 \beta^2 A} \left\{ In[\frac{m_0 c^2 \beta E}{2J^2(1-\beta^2)}] - [2(1-\beta^2)^{1/2} - 1 + \beta^2]\ln 2 + (1-\beta^2) + \frac{1}{8}[1-(1-\beta^2)^{1/2}]^2 - \delta_0 \right\} \quad (4)$$

$N_0$ is the Avogadro constant, $e$ is the electron charge, $Z$ is the atomic number, $A$ is the atomic weight, $m_0 c^2 = 0.511 MeV$ is the electron rest mass, $\beta = \{1-[m_0 c^2/(E+m_0 c^2)]^2\}$, $E$ is the kinetic energy of electron, $J(MeV)$ is the average ionization energy, $\delta_0$ is the density effect correction factor, and $\delta_0$ is introduced to reduce energy loss by electron collisions through target polarization.

The next step in electronic energy is decided by the law of energy logarithmic delay.

$$E_{n+1} = K E_n \quad (5)$$

K is 0.9576.

The continuous slow-down used to approximately calculate the interval mass range of electron.

$$\Delta S_{n+1} = \int_{E_{n+1}}^{E_n} \left|\frac{dE}{dS}\right|^{-1} dE \quad (6)$$

The electronic coordinates of step n+1 are as follows.

$$X_{n+1} = X_n + \Delta S_{n+1} \frac{\cos\theta_{n+1}}{\rho(j)} \quad (7)$$

where $\rho(j)$ is the lagrange interval density of j. $\theta_{n+1}$ is the positive angle between the electron motion direction and the X-axis of step n+1.

According to the Moliere Multiple Scattering Theory, the scattering angle is as follows.

$$F(\theta) = f^{(0)}(\theta) + f^{(1)}(\theta)/B + f^{(2)}(\theta)/B^2, \omega = X_c B^{1/2} \theta \quad (8)$$

where $B$, $Xc$ is one of energy parameters, $\theta$ is the discrete angle.

The electronic polar angle is as follows.

$$\theta_{n+1} = \cos^{-1}(\cos\theta_n \cos\omega + \sin\theta_n \sin\omega \cos\varphi) \quad (9)$$

where $\varphi$ is the electronic azimuth, $\varphi$ is the determined uniform sampling.

If the deposition energy of each incident electron is $\Delta E_i$ in the interval J, the total deposition energy of each incident electron is as follows.

$$E_n = \sum_{i=1}^{N} \Delta E_i \quad (10)$$

The specific energy of unit energy flux in interval J is as follow.

$$Q_J = \frac{E_n}{\rho_0 \Delta x N} \quad (11)$$

where $\Delta x$ is the mesh step.

According to the current and voltage waveform of experiment, the energy deposition of electron beam can be calculated, as shown as below:

$$E_R(x,t) = U_t I_t Q_J \Delta T / S \quad (12)$$

where $U_t$, $I_t$ are the instantaneous voltage and current of REB diode, $\Delta T$ is the time step, S is the spot area of electron beam on the target.

## 3. Comparison of Experimental Results and Numerical Calculation under Thermodynamic Parameters of Electron Beam Machine

The numerical calculation of the electron beam blow-off impulse and the thermal shock wave peak stress was carried out by using the software of thermodynamic effects of pulsed beam irradiation. By comparison, the calculation results were in good agreement with the experimental data. For the simulation experiment of electron beam irradiation, the electron beam is loaded according to the time course, with the electron beam diode current and voltage given by the experiment used as load conditions. With the voltage data corresponding to the electron beam energy, the energy deposition distribution of the electrons in the material can be obtained by the interpolation of the calculated energy deposition profile of the electron beam. With the current data corresponding to the intensity of the electron beam, the energy flux at that time can be obtained by combining the transmission efficiency of the electron beam in the drift tube with the beam spot area reaching the target and the beam energy. The experimental state of electron beam can be basically simulated, which is to the greatest extent consistent with the experimental process.

Fig. 2 shows the original REB current and voltage waveforms from the experiment. Due to the existence of many invalid data, the original data needs to be processed by extracting the valid pulse waveforms with negative current and voltage values in the same period. The valid data of REB current and voltage used in the calculation are then plotted in Fig. 2. Fig. 3 shows the electron beam energy spectrum – whose structure cannot be analysed because of the small grouping energy and whose data not directly used in the numerical calculation are included in current and voltage values – from the experiment through the processing of the current and voltage data. Meanwhile, this spectrum is the one after 0.01MeV grouping, from which the percentage of the electron beam in each energy range can be seen intuitively.

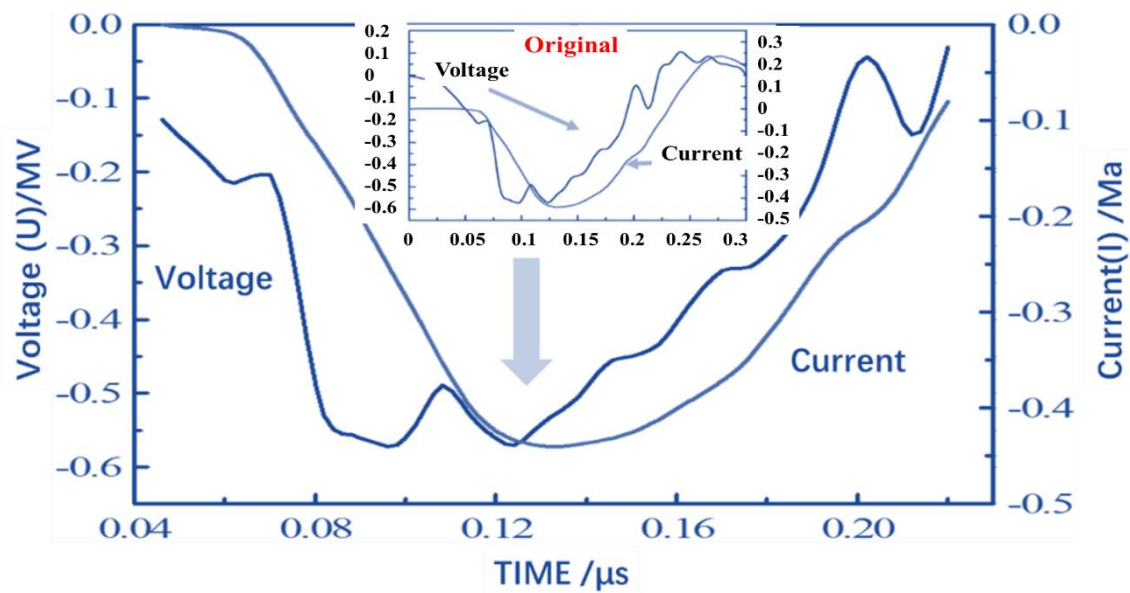

Fig. 2 Original REB current and voltage waveforms from the experiment

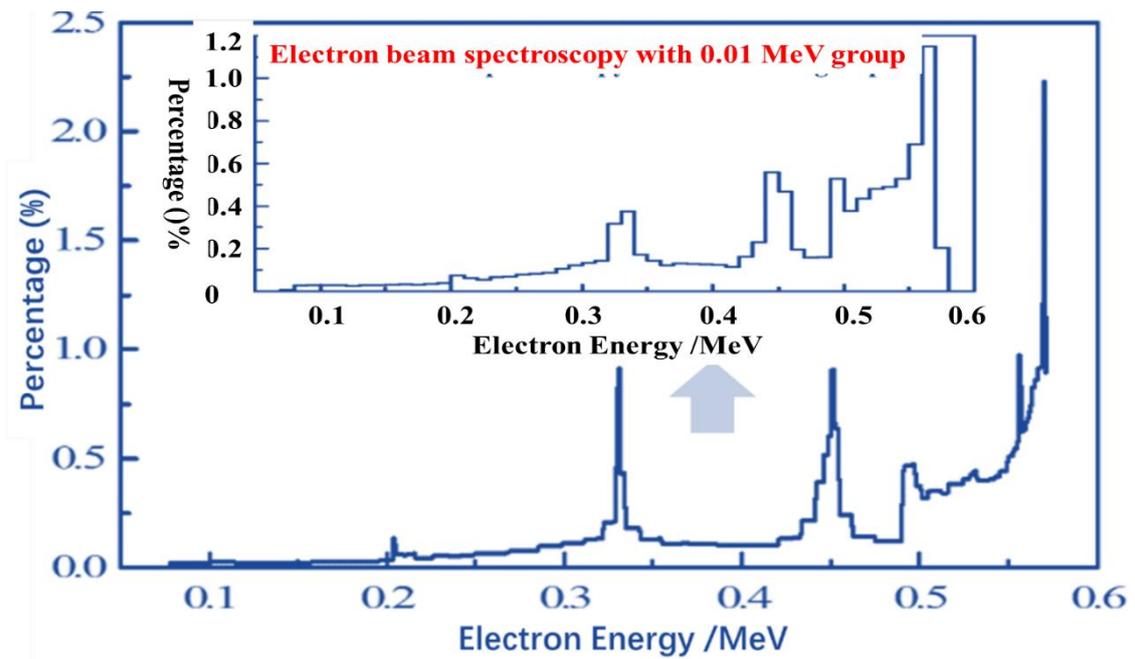

Fig. 3 Electron beam energy spectrum from the experiment after 0.01 MeV grouping

    According to the effective current and voltage data in the experiment, the transmission efficiency of the electron beam in the drift tube, and the beam spot area on the target surface, each experimental situation can be calculated by the finite difference method. Fig. 4 shows the comparison between the calculated results and the experimental data, from which it can be seen that the former are mostly in good agreement with the latter. For the larger ones, they are still in reasonable agreement with consideration of the uncertain electron beam energy flux. In a word, the former, which is credible, is consistent with the latter in the range of experimental uncertainty, and the numerical calculation is carried out well by using the numerical calculation software. Fig. 5 shows the comparison of the impulse coupling coefficients.

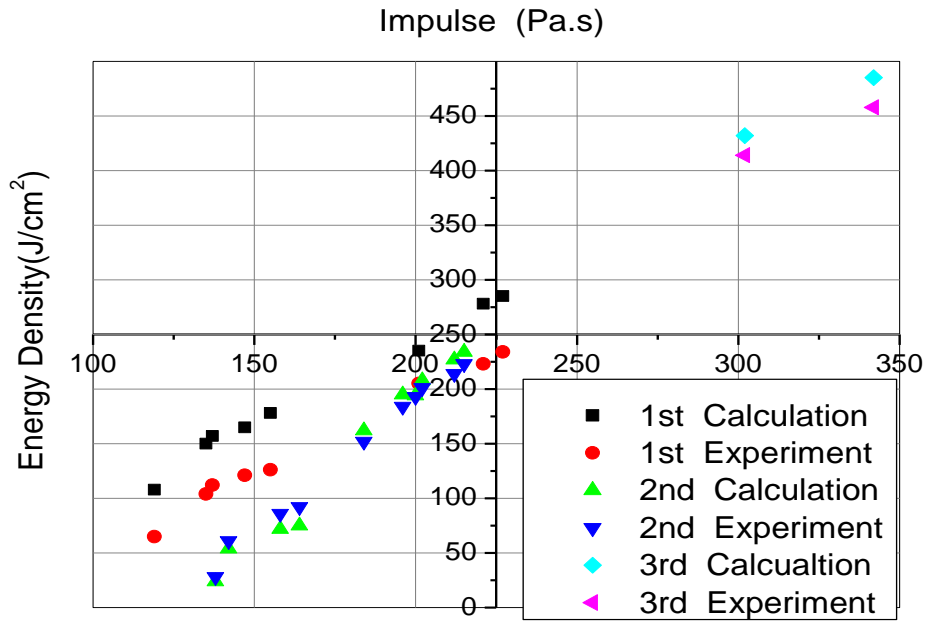

Fig. 4 Comparison between calculated results and experimental data of blow-off impulse

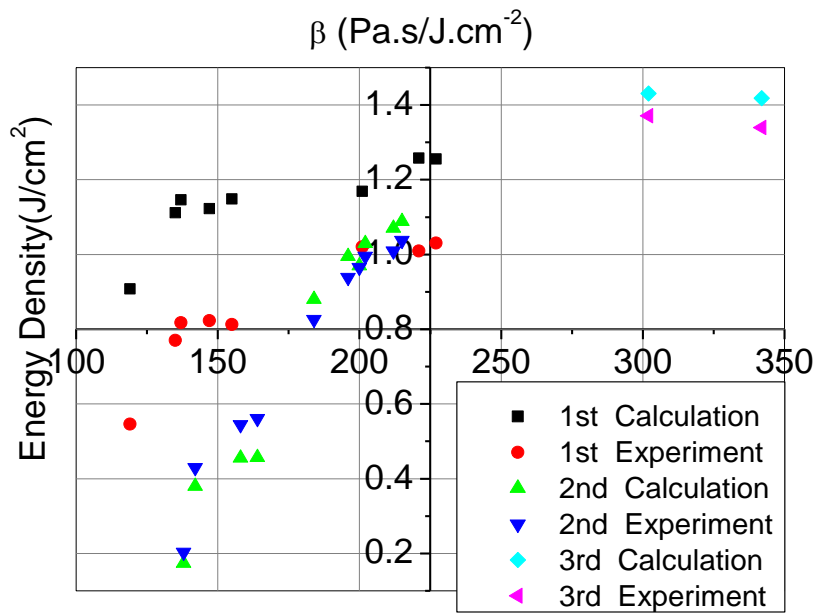

Fig. 5 Comparison between calculated results and experimental data of impulse coupling coefficient

## 4. High Energy Flux Expansion Calculation

The limited effective data of thermodynamic experiment and the small range of energy flux fail to meet the requirements of equivalence research. Therefore, this paper expands the experimental results and its range by means of numerical calculation. The energy flux range for equivalence research needs to be extended to 500J/cm² from 100-250J/cm² for the experiment. First, the calculation approach for energy flux in the electron beam irradiation experiment needs to be analysed.

$$\Phi = \frac{k\Delta t}{\pi R^2} \sum_i U(i)I(i) \quad (13)$$

where $U(i)$ and $I(i)$ represent the voltage and current data – including generally unchangeable system information of the electron beam accelerator – recorded in the experiment at each moment during the electron beam pulse irradiation. $\Delta t$ is the time interval of recording the current and voltage of the accelerator, and $k$ is the transmission efficiency of the electron beam in the drift tube. The failure of online measurement triggers the values of these two changing in a small range; $R$ is the radius of the electron beam spot on the target surface, with its size adjusted by changing the target position in the drift tube in the experiment. Moreover, different energy fluxes can be obtained by adjusting its size in the numerical calculation. This study gave out irradiation effects of the electron beam at high-energy flux by adjusting the beam spot radius and using numerical calculation for expansion. By taking the current and voltage as basic parameters and adjusting the beam spot radius, the blow-off impulse and thermal shock wave stress were experimentally calculated at the energy flux of 100-500J/cm², to complete the expansion of the experimental results of high-energy flux.

Fig. 6 shows the calculation results of blow-off impulse by experimental expansion, under the electron beam irradiation blow-off impulse with energy flux ranging 100-500J/cm², with maintained system information of electron beam. The numerical calculation results are equivalent to the experimental ones to a certain extent. The blow-off impulse increases linearly with the change of energy flux. Fig. 7 shows calculation results of blow-off impulse and impulse coupling coefficient under different energy fluxes

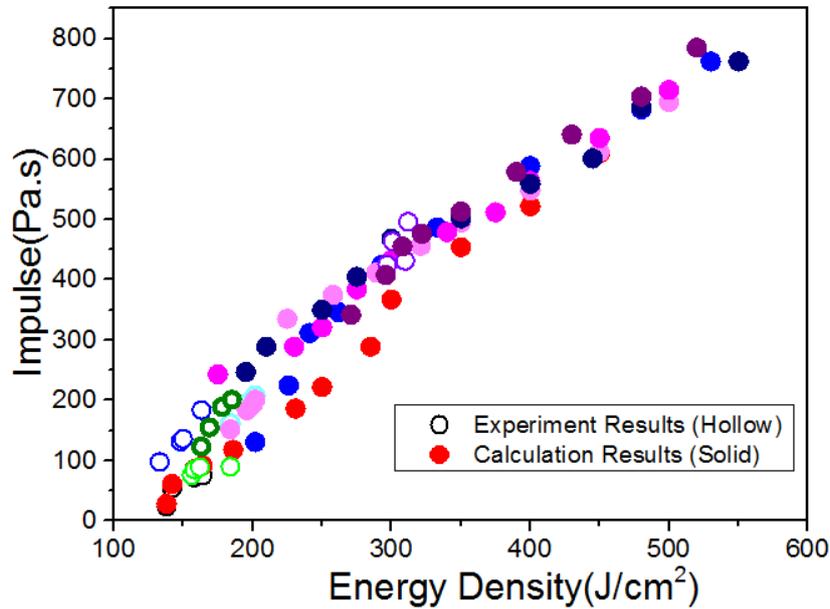

Fig.6 the calculation results of blow-off impulse by experimental expansion

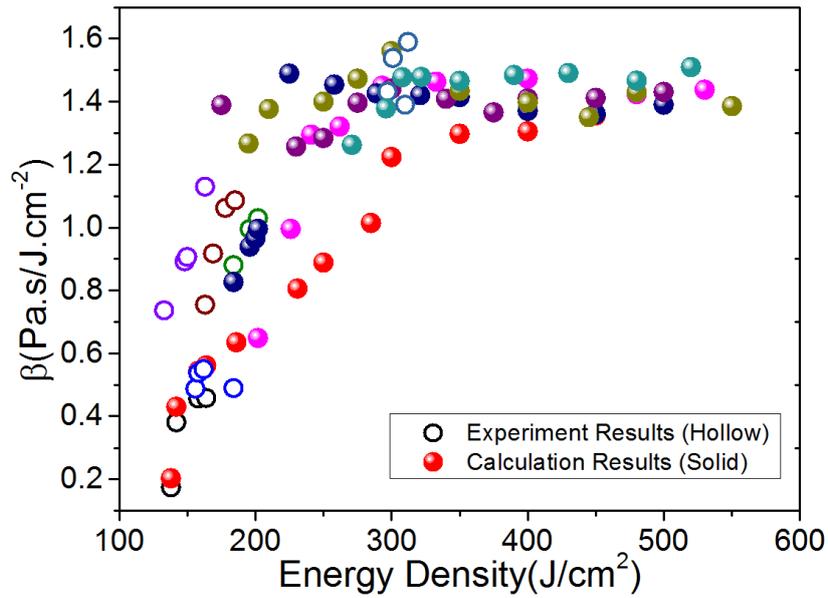

Fig. 7 Calculation results of blow-off impulse and impulse coupling coefficient under different energy fluxes

Fig. 8 shows the calculation results and experimental data of thermal shock wave stress under different energy fluxes. The former, albeit slightly larger, is in good agreement with the latter to a certain extent, with consideration of the uncertain energy flux and thermal shock wave stress. It shows the influence of uncertain electron beam system on the thermal shock wave stress, because of the linear increase of thermal shock stress with the change of the energy flux, and the small dispersion.

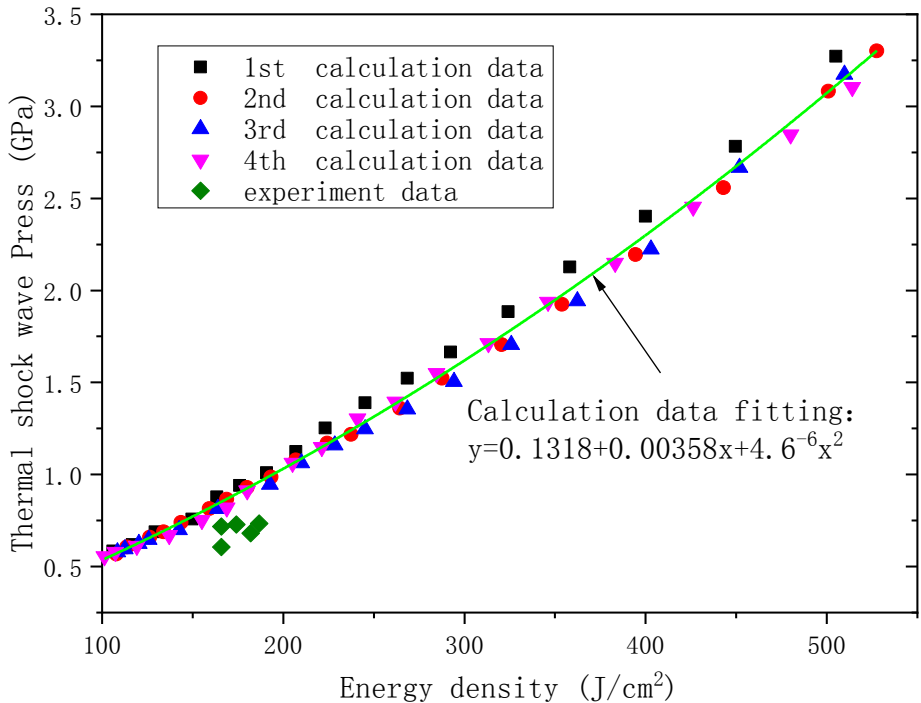

Fig. 8 Calculation and experimental results of thermal shock wave under different energy fluxes

To analyse the equivalence, the blow-off impulse results are calculated and compared with the experimental data, as shown in Fig. 9. Meanwhile, the data are fitted and the formulas are given.

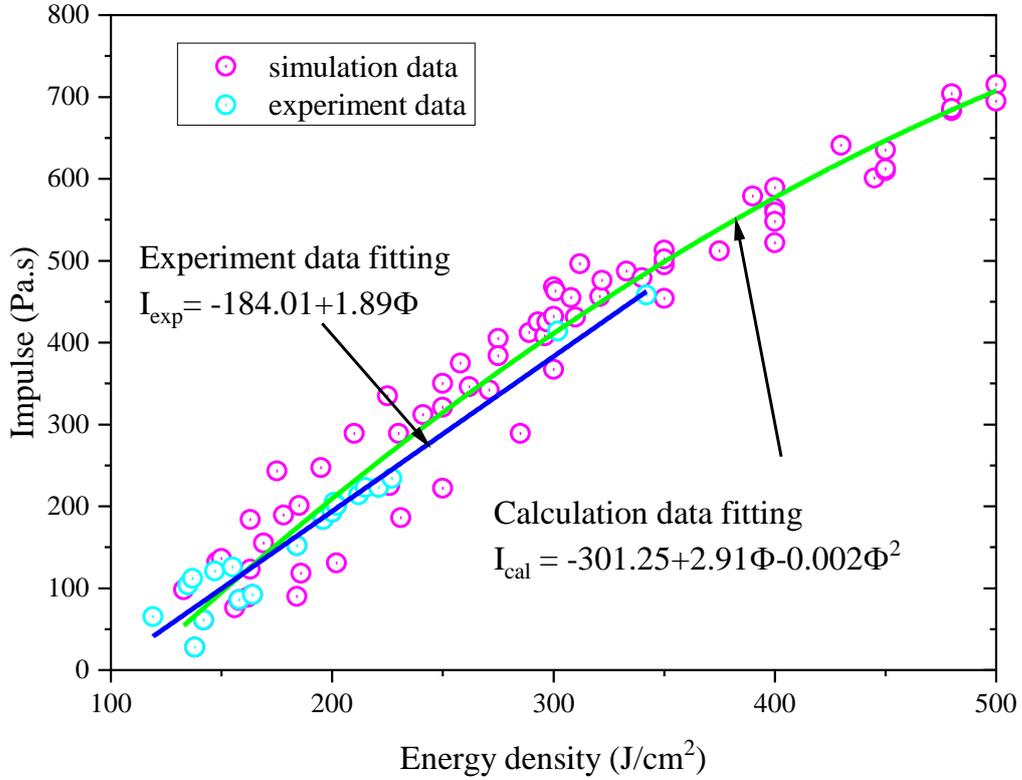

Fig. 9 Calculation and experimental results and fitting curves of electron beam irradiation blow-off impulse under different energy fluxes

## 5. Numerical Calculation of X-ray Blow-off Impulse

The same aluminium alloy materials were selected as the target materials in the experiment. The blackbody spectra were 1keV and 1.4 keV, and X-ray time spectrum used rectangular pulse of 0.1s in width. The variation of bow-off impulse with X-ray energy flux was calculated for X-ray spectrum, with its energy flux ranging 50-500J/cm$^2$. The calculation results of the blow-off impulse are shown in Fig. 10, with the black and red lines representing the blow-off impulse of 1keV and 1.4 keV blackbody spectrum X-rays respectively. It can be seen from the graph that the different monotonous increase of blow-off impulse with the energy flux. The approximate fitting formulas are as follows.

$$I_{kT=1.0} = 0.65\Phi_X + 2.0 \qquad (14a)$$

$$I_{kT=1.4} = 0.7\Phi_X - 25 \qquad (14b)$$

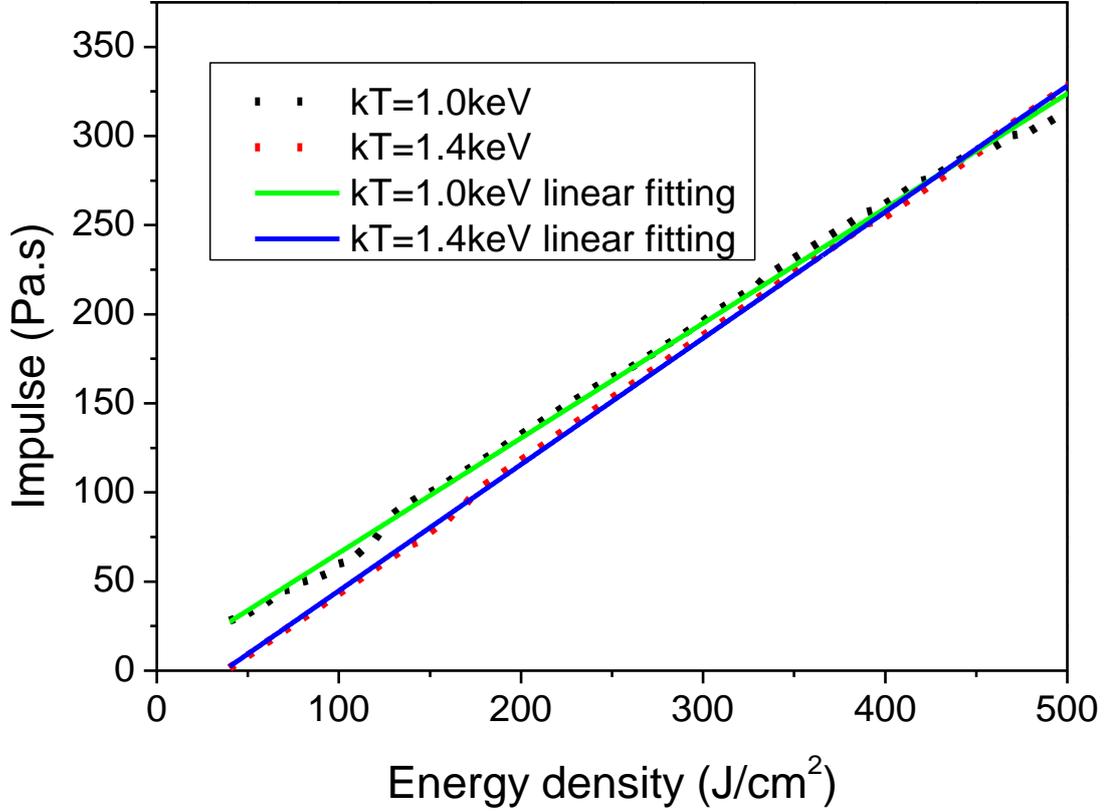

Fig. 10 Calculation result and fitting curve of blow-off impulse of blackbody spectrum X-ray under different energy fluxes

## 6. Equivalence Relations between Electron Beam and Blackbody Spectrum X-ray Irradiation Blow-off Impulse

According to the above calculation results of blow-off impulse, electron beam was adopted to simulate the blackbody spectrum X-ray irradiation blow-off impulse. After numerical analysis on the calculation results of both, it is found that there is a certain numerical correlation between the two blow-off impulses. Based on this, a simple numerical relation between the two blow-off impulses is as follows.

$$I_{kT=1.0} = 0.4 I_{electron} + 35 \ (Pa.s) \qquad (15a)$$

$$I_{kT=1.4} = 0.43 I_{electron} + 10 \ (Pa.s) \qquad (15b)$$

This formula gives out the numerical relations between the irradiation blow-off impulse of the electron beam and the blackbody spectrum X-ray under the same energy flux, which is called equivalent energy flux relations. Fig. 11 shows the comparison between the numerical calculation results and the blackbody spectrum X-ray's irradiation blow-off impulse converted from that of electron beam under identical energy flux relations. It can be seen intuitively that they are in good agreement.

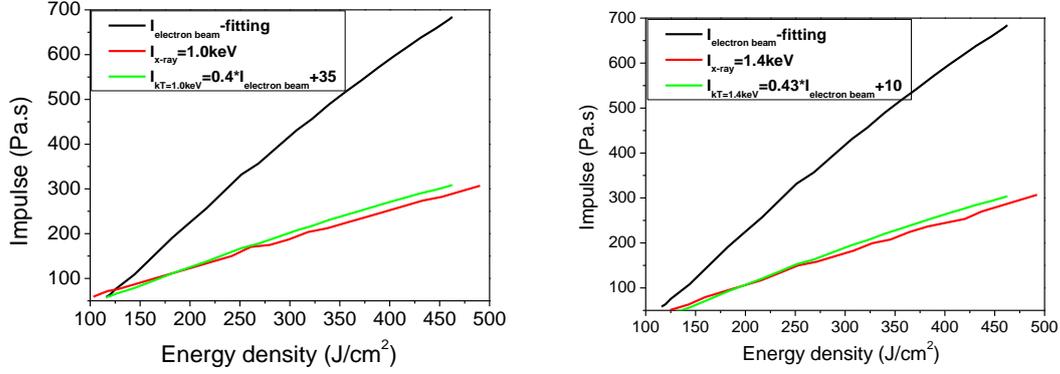

Fig. 11 Comparison between the results of energy-flux relation conversion and blackbody spectrum X-ray irradiation blow-off impulse

Formula (16) exhibits the numerical relation between electron beam and blackbody spectrum X-ray irradiation blow-off impulse at the same energy flux. However, it is not intuitive and convenient to directly apply it to equivalence and give out equivalence principles. Therefore, a further analysis of the relation between the data for the two blow-off impulses gave out the energy flux relation at the same blow-off impulse by the irradiation of electron beam and blackbody spectrum X-ray.

$$\Phi_{kT=1.0} = 2.5\Phi_{electron} - 130 \ (J/cm^2) \quad (16a)$$

$$\Phi_{kT=1.4} = 2.3\Phi_{electron} - 80 \ (J/cm^2) \quad (16b)$$

The above formula gives out the numerical relations of energy flux at the same blow-off impulse by the irradiation of electron beam and blackbody spectrum X-ray, which is called equivalent impulse relations. From the comparison between the numerical results and the blackbody spectrum X-ray's blow-off impulse converted from that of electron beam as per the above equivalent impulse relations, as shown in Fig. 12, it can be found that they are in good agreement.

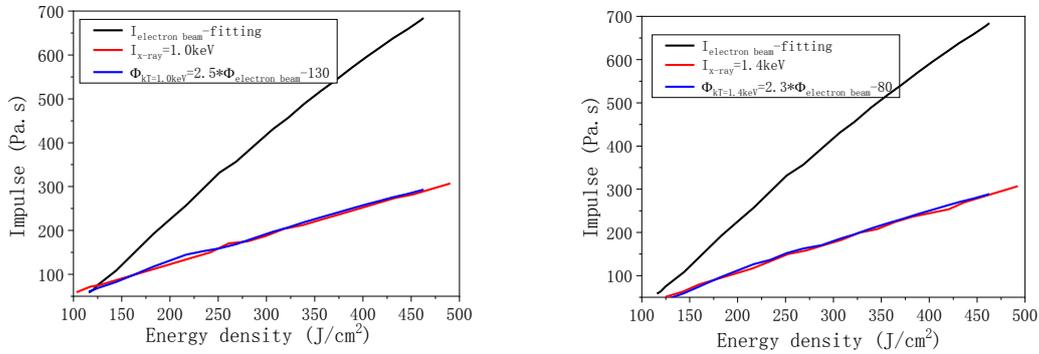

Fig. 12 Comparison between equivalent impulse-based conversion results and blackbody spectrum X-ray irradiation bow-off impulse

Fig. 13 shows the equivalent energy flux relations for measurement results of the experimental blow-off impulse. The results from comparison show that the simulation results of blackbody spectrum X-ray irradiation blow-off impulse converted from the experimental results of blow-off impulse as per the above flux relations are in great agreement with the numerical results. The experimental results of blackbody spectrum X-ray irradiation blow-off impulse at the same energy flux can be predicted according to the blow-off impulse by the electron beam experiment and the above flux relations. The experimental results of blackbody spectrum X-ray irradiation blow-off impulse could be indirectly obtained by the electron beam simulation experiment, in case of lack of the blackbody spectrum X-ray irradiation sources.

$$I_{kT=1.0} = 0.29 I_{electron(experimen)} + 68 \quad (Pa.s) \tag{17a}$$

$$I_{kT=1.4} = 0.3 I_{electron(experiment)} + 46 \quad (Pa.s) \tag{17b}$$

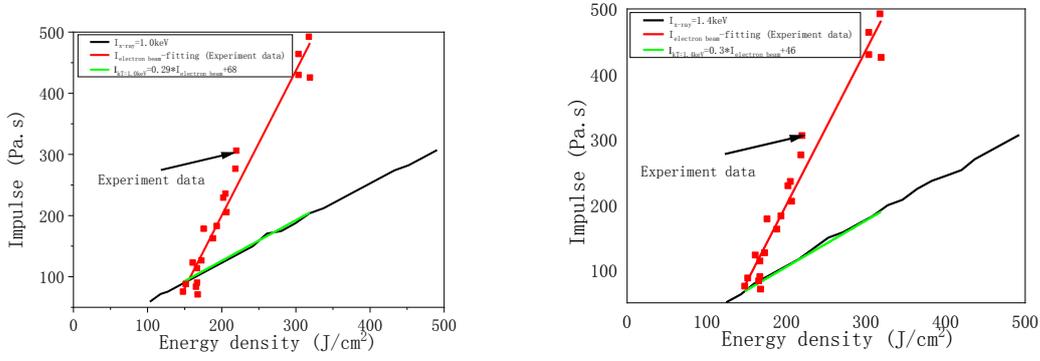

Fig. 13 Comparison between the fitting curves of experimental blow-off impulse after conversion via the equivalent energy flux relations and the blackbody spectrum X-ray irradiation blow-off impulse

Fig. 14 shows an equivalent impulse relation which is a direct equivalent one. The results obtained by numerical calculation are in good agreement with those obtained by X-ray irradiation. The electron beam experiment can be designed by using the equivalent impulse relations. The blackbody spectrum X-ray irradiation blow-off impulse value by direct measurement under the concerned flux is equivalent to the experimental result of blackbody spectrum X-ray irradiation blow-off impulse directly obtained by electron beam experiment without sufficient blackbody spectrum X-ray irradiation sources.

$$\Phi_{kT=1.0} = 3.75 \Phi_{electron(experiment)} - 425 \quad (J/cm^2) \tag{18a}$$

$$\Phi_{kT=1.4} = 3.38 \Phi_{electron(experiment)} - 340 \quad (J/cm^2) \tag{18b}$$

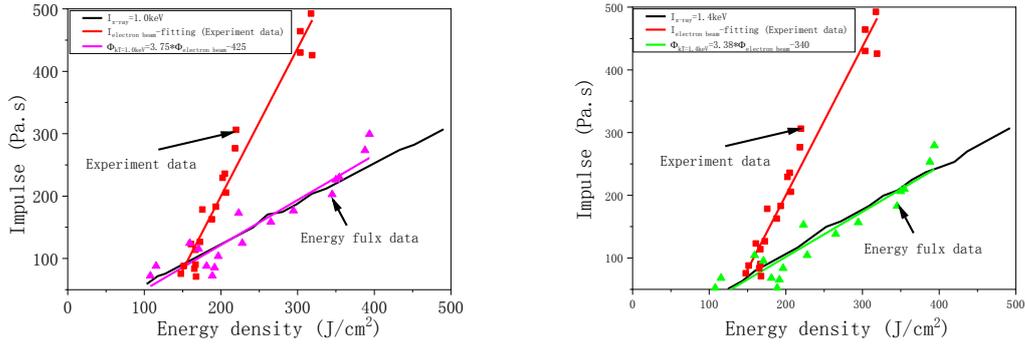

Fig. 14 Comparison between the fitting curves of experimental blow-off impulse after conversion via equivalent impulse relations and the blackbody X-ray irradiation blow-off impulse

## 7. Conclusions

This paper studied the equivalence problems in thermodynamics effects between the multi-energy composite spectrum electron beam and blackbody spectrum X-ray, which could provide a reference for rapid determination of experimental parameters in simulation experiments.

(1) The electronic beam experimental platform and impulse measurement system were introduced to analyse the set input parameters by numerical calculation on the basis of experimental conditions. By comparison of the impulse, impulse coupling coefficient, it was shown that the credible calculation results were consistent with the experimental data in the range of experimental uncertainty. The numerical simulation was carried out well by using the numerical calculation software.

(2) Experiment results and energy fluence range were expanded by numerical calculation to meet the requirements of equivalence study, and the data for electron beam irradiation blow-off impulse with energy fluence of 100-500J/cm$^2$ were thus obtained. This result meanwhile maintained the input information of electron beam system. From impulse calculation results, the blow-off impulse presented a linear increase with energy fluence.

(3) The RAM software with a self-developed programme was employed to achieve the X-ray irradiation blow-off impulse and the approximate fitting formula under blackbody spectrum of 1keV and 1.4keV. For input conditions, the aluminium alloy material as same as the target material for the experimental measurement was selected, while the rectangular pulse of 0.1μs in width was adopted for X-ray time spectrum, thus obtaining the energy fluence of 50-500J/cm2.

(4) By comparison between the results of electron beam experiment, simulation and the blackbody spectrum X-ray numerical simulation, the equivalent flux and impulse relations were obtained.

In a word, based on the actual results of the electron beam simulation experiment, the results were also expanded by numerical calculation. The numerical calculation of the blackbody spectrum X-ray irradiation blow-off impulse gave out the variation curve of the X-ray blow-off impulse with the energy flux. Based on this, the equivalent flux and impulse relations were obtained by analysing the numerical and experimental calculation results of the electron beam blow-off impulse and the calculation results of the X-ray blow-off impulse. Through such relations, the experimental results of blackbody spectrum X-ray irradiation blow-off impulse could be obtained directly or indirectly by the electron beam simulation experiment. This is of certain practical significance.

**Funding Statement**

This work was supported by the key project of Intergovernmental International Scientific and Technological Innovation Cooperation in China under Grant No.2016YFE0128900, and the National Natural Science Foundation of China under Grant No.11775166.

**Data Availability**

The data used to support the findings of this study are available from the corresponding author upon request.

**Conflicts of Interest**

The authors declare that they have no conflicts of interest.